\documentclass[journal=jpclcd,manuscript=letter]{achemso}

\usepackage{amsmath,multirow,amssymb,graphicx,rotating,pinlabel}
\usepackage{graphics,graphicx,amsmath}
\usepackage[usenames]{color}
\newcommand{\COMMENT}[1]{}
\usepackage{pdfsync}

\date{\today}
\author{Thomas~van~Dijk}
\affiliation{Beckman Institute for Advanced Science and Technology, University of Illinois at Urbana-Champaign, Urbana, IL 61801 USA}
\author{Sean~T.~Sivapalan}
\affiliation{Department of Materials Science and Engineering, University of Illinois at Urbana-Champaign, Urbana, IL 61801 USA}
\author{Brent~M.~DeVetter}
\affiliation{Department of Electrical and Computer Engineering, University of Illinois at Urbana-Champaign, Urbana, IL 61801 USA}
\author{Timothy~K.~Yang}
\affiliation{Department of Chemistry, University of Illinois at Urbana-Champaign, Urbana, IL 61801 USA}
\author{Matthew~V.~Schulmerich}
\affiliation{Beckman Institute for Advanced Science and Technology, University of Illinois at Urbana-Champaign, Urbana, IL 61801 USA}
\alsoaffiliation{Department of Bioengineering, University of Illinois at Urbana-Champaign, Urbana, IL 61801 USA}
\author{Catherine~J.~Murphy} 
\affiliation{Department of Materials Science and Engineering, University of Illinois at Urbana-Champaign, Urbana, IL 61801 USA}
\alsoaffiliation{Department of Chemistry, University of Illinois at Urbana-Champaign, Urbana, IL 61801 USA}
\author{Rohit~Bhargava} 
\affiliation{Beckman Institute for Advanced Science and Technology, University of Illinois at Urbana-Champaign, Urbana, IL 61801 USA}
\alsoaffiliation{Department of Bioengineering, University of Illinois at Urbana-Champaign, Urbana, IL 61801 USA}
\alsoaffiliation{Department of Mechanical Science and Engineering, Chemical and Biomolecular Engineering and University of
   Illinois Cancer Center, University of Illinois at Urbana-Champaign, Urbana, IL 61801 USA}
\author{P.~Scott~Carney}
\email{carney@uiuc.edu} 
\affiliation{Beckman Institute for Advanced Science and Technology, University of Illinois at Urbana-Champaign, Urbana, IL 61801 USA}
\alsoaffiliation{Department of Electrical and Computer Engineering, University of Illinois at Urbana-Champaign, Urbana, IL 61801 USA}

\title[]{Competition between extinction and enhancement in surface enhanced Raman spectroscopy}

\begin{document}
 
\begin{abstract}
Conjugated metallic nanoparticles are a promising means to  achieve ultrasensitive and multiplexed sensing in intact three-dimensional samples, especially for biological applications,  via surface enhanced Raman scattering (SERS).
We show that  enhancement and extinction are linked and compete in a collection of metallic nanoparticles.    Counterintuitively, the Raman signal vanishes when nanoparticles are excited at their plasmon resonance, while increasing nanoparticle concentrations at off-resonance excitation sometimes leads to decreased signal. We develop an effective medium theory that explains both phenomena. Optimal choices of excitation wavelength, individual particle enhancement factor and concentrations are indicated.  
\end{abstract}

\begin{figure}[ht]
\includegraphics[width=3.25in]{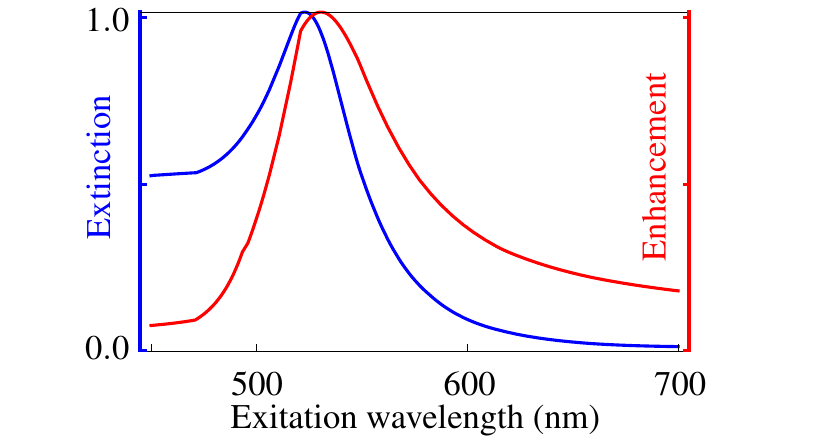}
\caption{In blue: The normalized extinction cross-section $C_{\mathrm{ext}}$ from Eq.~\ref{eq:abs}. Extinction by gold spheres of $5$ nm radius in aqueous suspension as a function of the wavelength of the incident light.
         In red: The normalized Raman enhancement, $G(\lambda)$ Eq.~\ref{eq:enhance} vs excitation wavelength for gold spheres of radius much smaller than the wavelength evaluated for a Raman shift of $0$ nm.}
\label{fig1}
\end{figure}

Several methods for using surface-enhanced Raman scattering (SERS)\cite{schatz_electromagnetic_2002} have emerged for biomedical applications ultrasensitive sensing and multiplexed analyses. In particular, nanoparticles have been the focus of recent efforts towards in vitro and in vivo molecular sensing \cite{lyandres_progress_2008,qian_vivo_2008,von_maltzahn_sers-coded_2009}. Nanoparticles can dramatically increase the electric field intensity near and at their surface, providing useful SERS-based probes, especially for deep tissue imaging at varying concentrations \cite{stone_prospects_2010}. Typically, a nanostructured particle is bioconjugated and employed in the same manner that conventional fluorescent probes are used for molecular imaging. SERS probes are postulated to offer bright and stable signals and extensive multiplexing \cite{cao_raman_2003}, while it has been assumed that experimental best practice parallels that of fluorescent probes, i.e. that one should  excite at the strongest resonance and use a high concentration. Thusfar, the design of nanoparticle-based SERS experiments has focused on maximizing the local electromagnetic field enhancement in or around an individual particle \cite{talley_surface-enhanced_2005,kodali_optimally_2010}. This strategy fails to take into account the physics of propagation in the bulk medium where the same processes which give rise to enhancement also lead to increased extinction of both the illumination and the Raman scattered light. Particles provide enhanced fields for Raman scattering and the same particles form an effective medium with corresponding absorption. The importance of absortion of the Raman scattered light is recognized in \cite{jackson_controlling_2003}. However, they do not describe the necessary link and competition between the enhancement and the extinction. For example, it is commonly known to experimentalists that gold nanospheres exhibit a plasmon resonance at 520 nm and should produce a large local field enhancement when illuminated at 532 nm; yet, no appreciable Raman signal is observed upon 532 nm excitation commonly ascribed to interband transitions in gold \cite{alvarez-puebla_effects_2012}. Away from the plasmon resonance frequency maximum, the Raman signal is again observed and actually increases as the excitation wavelength becomes longer. 

In this Letter, we address the issue of extinction by a suspension of nanoparticles in SERS experiments  through an effective-medium approach.  It is shown that extinction and enhancement are tied to each other and compete in such a way that peak signals are acquired off resonance and that, at any wavelength, an optimal particle concentration exists to maximize the Raman signal.   We provide verification of the model with experiments in which the particle concentration is varied. 

Propagation of light in a dilute suspension of identical particles is well-approximated by propagation through a homogeneous medium with an effective refractive index $\widetilde{m}$, given by \cite{bohren_applicability_1986}
\begin{equation}
\widetilde{m}=m\left[1+\mathrm{i}\frac{2\pi\rho}{k^{3}}S(0)\right],
\label{index}
\end{equation}
where $m$ is the refractive index of the medium in which the particles are embedded, $k=\omega/c$ is the wavenumber in the medium, $\rho$ is the number of particles per unit volume and $S(0)$ is the scattering amplitude in the forward direction \cite{bohren_absorption_1983}. The absorption coefficient in a medium with a complex refractive index is $\alpha=2 k\,\mathrm{Im} \,\tilde m$. For a suspension with small identical particles the absorption coefficient is given by\,
$\alpha={m 4\pi\rho}{k^{-2}}\mathrm{Re}\left[S(0)\right]=\rho m  C_{\mathrm{ext}}$,
where $C_{\mathrm{ext}}$ is the extinction cross section of a single particle in the suspension, proportional to the real part of the forward scattering amplitude. The attenuation of a well-collimated beam propagating through the effective medium is described by Beer's law \cite{van_de_hulst_light_1982}, $I(h)=I(0)e^{-h m\rho C_{ext}}$ where $I$ is the field intensity, and $h$ is the propagation distance.

The extinction cross-section, $C_{\mathrm{ext}}$, for a small metallic sphere with radius $a$, to terms of order $(k a)^{4}$, is given by \cite{bohren_absorption_1983}
\begin{equation}
\begin{split}
&C_{\mathrm{ext}}=4k\pi a^{3}\mathrm{Im}\left\{\frac{p^{2}-1}{p^{2}+2}\left[1+\frac{(k a)^{2}}{15}\left(\frac{p^{2}-1}{p^{2}+2}\right)\times\right.\right.\\
&\left.\left.\frac{p^{2}+27 p^{2}+38}{2 p^{2}+3}\right]\right\}+\frac{8}{3}(ka)^{4}\pi a^{2}\mathrm{Re}\left[\left(\frac{p^{2}-1}{p^{2}+2}\right)^{2}\right],
\end{split}
\label{eq:abs}
\end{equation}
where $p=m_s/m$ is the ratio of the refractive index of the material of the spheres, $m_s$, to that of the refractive index of the medium, $m$, which both depend on the wavenumber. For dilute suspensions, the change in  real part of the refractive index of the effective medium from the background is negligible. The extinction from gold spheres in a suspension is shown in Figure\ref{fig1}(a) where the extinction peaks near the Fr\"{o}hlich frequency ($\lambda_{f}\approx 520$nm), for this calculation the optical constants obtained by Johnson and Christy for gold have been used \cite{johnson_optical-constants_1972}.

The Raman signal, which we denote $R$, from a single, isolated nanoparticle, depends on the incident field amplitude, $E_{0}$, the number of Raman-active molecules, $N$, the local field enhancement, $f(\boldsymbol{r},\omega)$, and the spatial distribution of those molecules.  This last point we address through a probability density, which in general will also depend on the number of molecules present,  $p(\boldsymbol{r},N)$.  Though not explicitly noted, the local enhancement factor is also dependent on the orientation of the incident electric field vector.  The number of molecules attached to the nanoparticle may itself be random and given by probability of finding $N$ molecules attached to the particle $P_N$.  A single molecule at $\boldsymbol{r}$ is excited by a field with amplitude $E_0f(\boldsymbol{r},\omega_0)$ producing a secondary source proportional to the Raman susceptibility $\chi$, which implicitly depends on $\omega_0$, and $\omega$.  The field reradiated at the Raman-shifted frequency $\omega$ is enhanced by the particle as well so that, by reciprocity, the reradiated field is proportional to $\chi E_0f(\boldsymbol{r},\omega_0)f(\boldsymbol{r},\omega)$.  We assume the Raman signal from each reporter molecule is statistically independent, so the intensities add. 
The ensemble-averaged Raman signal for a single nanoparticle is thus given by
\begin{align}
 R&=|\chi|^2\sum_{N=1}^{\infty} NP_N\int {\mathrm d}^3r \left|E_0f(\boldsymbol{r},\omega_0)f(\boldsymbol{r},\omega)\right|^{2}p(\boldsymbol{r}, N)\nonumber\\
 &=\left< N\right>G R^{(0)},
\label{eq:Ramansingle}
\end{align}
where $R^{(0)}$ is the Raman signal from one molecule absent the particle, and $G$ is the Raman enhancement factor and generally depends on $p(\boldsymbol{r}, N)$ and $P_N$. For systems in which the particle placement is independent of the number of particles, the sum and the integral may be carried out independently, the sum yielding the average number of molecules $\left<N\right>$, and the integral resulting in a $G$ independent of the number of molecules.

The enhancement factor for a small sphere of radius $a$,  $(a\ll\lambda)$,  with a uniform probability of molecule placement over the surface of the sphere,  can be calculated in closed form \cite{kerker_surface_1980},
\begin{equation}
G(\omega,\omega_{0})=|[1+2g(\omega_{0})][1+2g(\omega)]|^{2},
\label{eq:enhance}
\end{equation}
where $g=(p^{2}-1)/(p^{2}+2)$, $\omega_{0}$ is the frequency of the incident field and $\omega$ is the frequency of the Raman-scattered field.

\begin{figure}[ht]
\includegraphics[width=3.25in]{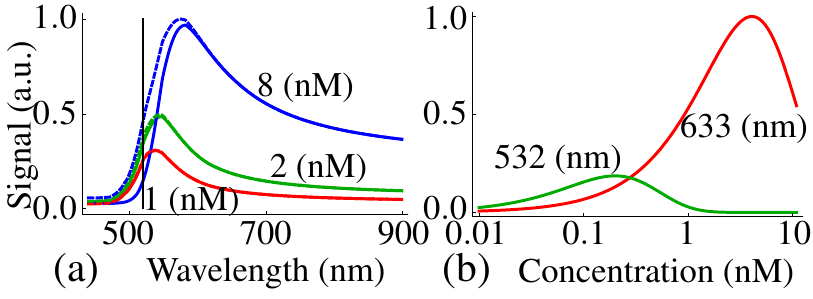}
\caption{(a) Solid lines: Predicted signal in transmission mode vs wavelength of the incident light.
 Dashed lines: Predicted signal in reflection mode vs wavelength of the incident light.  Both transmission and reflection signals are with three different concentrations of the nanospheres which have a radius 6 nm, the sample thickness h is 2 cm. The vertical black line indicates the location of the surface plasmon resonance.
(b) Predicted signal in transmission mode vs concentration, for two different incident wavelengths, the radius of the spheres is 15  $\rm{nm}$.}
\label{fig2}
\end{figure}

The enhancement calculated by Eq.~\ref{eq:enhance}, is shown in Figure~\ref{fig1} along side the extinction using the optical constants obtained by Johnson and Christy\cite{johnson_optical-constants_1972}. It is clear that enhancement and extinction are closely linked and that when the enhancement is strong, the correspondingly strong extinction must be taken into account.   
The light falling on a single particle is attenuated  by propagation through the suspension and arrives with amplitude attenuated by the factor  $\exp\left[-\int_0^z {\mathrm d}z' \rho(z') mC_{\mathrm{ext}}(\omega_{0})/2\right]$.  
The local Raman signal is then $\left< N\right>R^{(0)}G\rho(z)\exp\left[-\int_0^z {\mathrm d}z' \rho(z') mC_{\mathrm{ext}}(\omega_{0})\right]$.  
In transmission mode, this signal must then propagate out through the medium to $z=h$ and the intensity is attenuated by a factor $\exp\left[-\int_z^h {\mathrm d}z' \rho(z') mC_{\mathrm{ext}}(\omega)\right]$.  
The total signal is a sum over the signal from all particles so that 
\begin{align}
R&= \left< N\right>AR^{(0)} G\int_{0}^{h}\mathrm{d}z\,\rho(z)\nonumber \\
&\times\exp\left[-\int_0^z {\mathrm d}z' \rho(z') mC_{\mathrm{ext}}(\omega_{0})\right]\nonumber\\
&\times\exp\left[-\int_z^h {\mathrm d}z' \rho(z') mC_{\mathrm{ext}}(\omega)\right],
\label{eq:transmission}
\end{align}
where $A$ is the integral over the transverse beam profile normalized to peak value, the effective transverse area of the beam.
When the concentration $\rho(z)$ does not depend on $z$, the integrals can be computed in closed form with the result
\begin{equation}
R= \left< N\right> AR^{(0)} G\frac{e^{- m C_{\mathrm{ext}}(\omega_{0})h\rho}-e^{-m C_{\mathrm{ext}}(\omega)h\rho}}{m C_{\mathrm{ext}}(\omega)- m C_{\mathrm{ext}}(\omega_{0})}.
\label{eq:analtrans}
\end{equation}
From this expression it is seen that there are two competing processes that determine the size of the Raman signal: The enhancement, $G$, and the extinction that results in a exponential decay of the signal. The same processes that increase the enhancement also increase the extinction. The attenuation due to extinction depends not only on the frequency, but also on the concentration of the nanospheres.  This is illustrated in Figure~\ref{fig2}(a), where it is shown that for increasing concentration, the peak of the signal is shifted farther away from the resonant wavelength. This result explains the absence of Raman signal at the plasmon resonance where extinction is so strong that no signal is observed. 

In reflection mode there is always a contribution from the front layer of the sample which is not attenuated and so the expression for the Raman signal is slightly altered,
\begin{equation}
R= \left< N \right>AR^{(0)}  G\frac{1-e^{-hm\rho\left[C_{\mathrm{ext}}(\omega)+C_{\mathrm{ext}}(\omega_{0})\right]}}{mC_{\mathrm{ext}}(\omega)+mC_{\mathrm{ext}}(\omega_{0})}.
\label{eq:reflection}
\end{equation}
The Raman signal in reflection mode for three different concentrations of the nanospheres in shown in Figure~\ref{fig2}(a) as the dashed lines. In the reflection mode there is a slightly higher signal to the blue side of the resonance compared to the signal in transmission mode.

The Raman signal in transmission mode is depicted in Figure~\ref{fig2}(b) for two commonly used wavelengths evaluated for a Raman band at 1076 $\mathrm{cm}^{-1}$. For $\lambda=532$ nm, the excitation wavelength closest to the plasmon resonance, the signal is very small. A higher signal is found farther away from resonance with the peak shifted to the red. For relatively low concentrations, the biggest signal is obtained with a wavelength of $632$ nm. Only for concentrations smaller than $0.1$ nM the signal is bigger for excitation wavelength closest to resonance, as is shown in Fig~\ref{fig2}(b). It is seen that there is an concentration that maximizes the signal. This optimal concentration, $\rho_{\mathrm{opt}}$, can be found by differentiating Eq.~\eqref{eq:analtrans} and equating it to zero, giving the following expression
\begin{equation}
\rho_{\mathrm{opt}}=\frac{\ln\left[C_{\mathrm{ext}}(\omega)/C_{\mathrm{ext}}(\omega_{0})\right]}{h m\left[C_{\mathrm{ext}}(\omega)-C_{\mathrm{ext}}(\omega_{0})\right]}.
\label{eq:optimalconc}
\end{equation}
When the extinction cross-section, $C_{\mathrm{ext}}(\omega)$, equals, or is very close to, $C_{\mathrm{ext}}(\omega_{0})$, the optimal concentration becomes $\rho_{\mathrm{opt}}=1/\left[h m C_{\mathrm{ext}}(\omega_{0})\right]$.
The strong nonlinearity with concentration that these competing phenomena impose on the recorded signal is also a caution in the development of practical assays and must be taken into account to correctly quantify results across samples. Hence this physics-based analysis enables quantitative molecular imaging for SERS-based microscopy.

\begin{figure}[ht]
\includegraphics[width=3.25in]{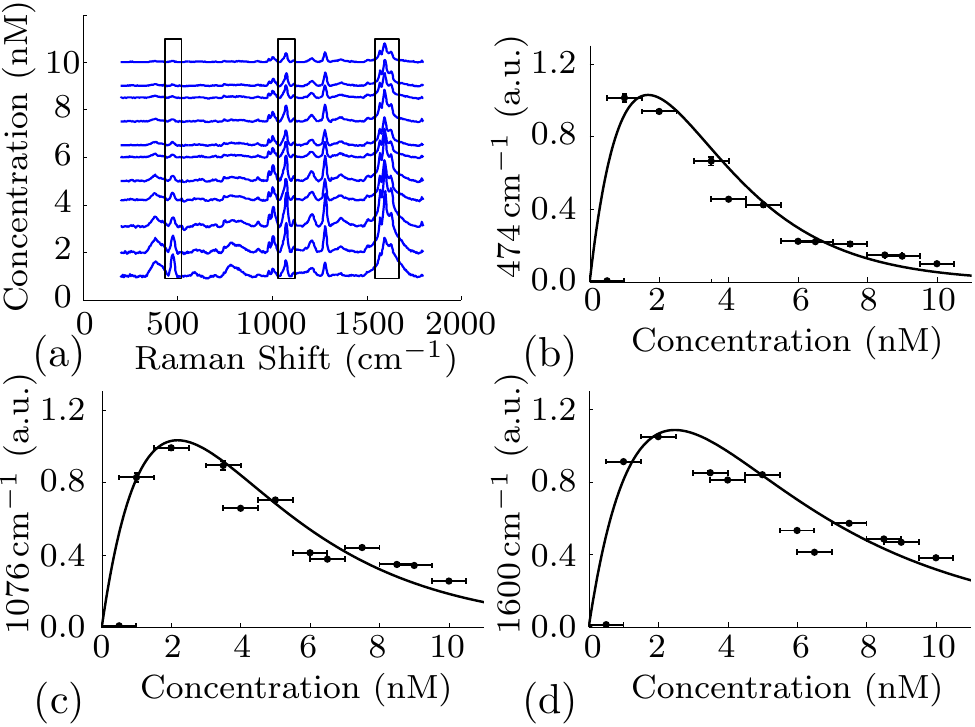}
\caption{(a) SERS spectra of $4,4'-$dipyridyl attached to gold nanospheres with a radius of 15 nm at an excitation wavelength of 633 nm at different nanoparticle concentrations. Measured Raman signals (points) agree with the theoretical prediction (solid line) for a Raman shift of (b) 474 $\mathrm{cm}^{-1}$, (c) 1076 $\mathrm{cm}^{-1}$ and (d) 1600 $\mathrm{cm}^{-1}$.}
\label{fig7}
\end{figure}

The model presented in this paper is validated by measuring the SERS  signal of 4,4'-Dipyridyl Raman reporter molecules attached to gold nanospheres. Spectra were acquired from the nanoparticles in suspension using a high-resolution Raman spectrometer (LabRAM, Horiba) with a 90 second acquisition time.  The Raman shift from 200 to 1800 cm$^{-1}$ was collected at 10 cm$^{-1}$ resolution with 10 mW laser power at the sample.  Transmission Raman measurements were collected by focusing laser light through a 1 cm  cuvette with a 50 mm focal-length lens and collected with a 100 mm focal-length lens to collimate the transmitted light and direct it to the spectrograph.

The integrated SERS signal under three different bands (476 $\mathrm{cm}^{-1}$, 1076 $\mathrm{cm}^{-1}$, 1600 $\mathrm{cm}^{-1}$) is compared for different concentrations of the gold spheres when excited at $632$ nm. The SERS spectra from 4,4'-Dipyridyl for increasing concentrations is illustrated in Figure~\ref{fig7}(a). The three boxes indicate the Raman bands for which the signal is investigated as a function of concentration. The signal is obtained by integrating the Raman band of interest over the width of the box as shown in Figure~\ref{fig7}. 

As predicted, increasing the concentration of nanoparticles in an attempt to increase the signal leads to signal attenuation beyond an optimal concentration.  The measurements are in good agreement with the model. Our results suggest that strategies to increase Raman signals using nanoparticles should not focus on achieving greater local enhancement but instead might strive for designs that maximize total signal by separating the single-particle enhancement and absorption peaks or otherwise tailoring shape of the enhancement and absorption curves to maximize the gap between absorption and enhancement at frequencies away from resonance.  A move towards using thin samples with large areas of collection is also suggested.  We see that signal is increased by moving away from resonance and, in some cases, by lowering the concentration of particles.  While we focused on nanospheres, our results apply broadly to particle-based Raman enhancement with nonspherical particles as well.  

\section{Experimental methods}
Gold nanospheres of 15 nm radius were synthesized by the boiling citrate method \cite{ji_size_2007,kimling_turkevich_2006}. For stability against aggregation, 100 mg of bis(p-sulfonatophenyl)phenylphosphine dihydrate dipotassium salt (BSPP) was added to 100 ml of as-synthesized nanoparticles \cite{loweth_dna-based_1999,schmid_complexation_1989}. The mixture was left to stir overnight (12-16 hours) and excess reagents removed by two centrifugation cycles (3000 RCF, 20 mins). For 4,4'-dipyridyl complexation, 1 ml of 10 mM  4,4'-dipyridyl in water was added to 9 ml of BSPP stabilized gold nanoparticles and left to complex overnight \cite{lim_highly_2011}. Excess reagents were removed by two centrifugation cycles (3000 RCF for 20 mins). For final purification, we dialyzed the solutions in Thermo Scienfific G2 Slide- A- Lyzer G2 cassettes against $4 L$ of Barnstead E-Pure ($18 \mathrm{M}\Omega\mathrm{cm}$) water for 48 hours.

\acknowledgement
The work was supported in part by the Beckman fellows program. S.T.S. and B.M.D. acknowledge support from the University of Illinois at Urbana-Champaign from NIH National Cancer Institute Alliance for Nanotechnology in Cancer ``Midwest Cancer Nanotechnology Training Cente'' Grant R25CA154015A. M.V.S. acknowledges support through the Congressionally Directed Medical Research Program Postdoctoral Fellowship BC101112. We also acknowledge support from a Beckman Institute seed grant, AFOSR Grant No.FA9550-09-1-0246 and NSF Grant Nos. CHE-1011980 and CHE 0957849.

%\bibliography{colloids}
%\begin{thebibliography}{999}
%\end{thebibliography}
\providecommand*\mcitethebibliography{\thebibliography}
\csname @ifundefined\endcsname{endmcitethebibliography}
  {\let\endmcitethebibliography\endthebibliography}{}

\end{document}